\documentclass[amsmath,amssymb,prb,showpacs,reprint]{revtex4-1}

\usepackage{graphicx,color}

\usepackage{gensymb} %\micro, \ohm, \degree, \celsius
\usepackage[seperr, load={}]{siunitx}
\usepackage{ulem} % strikethru command

\newcommand{\figref}[1]{Fig.~\ref{#1}}
\def\paper{Article}

\DeclareMathOperator{\tr}{Tr}
\DeclareMathOperator*{\Real}{\Re{\rm e}}

\newcommand{\nn}{\nonumber}
\newcommand{\ket}[1]{\left|{#1}\right\rangle}

\newcommand{\braket}[2]{\langle#1|#2\rangle}
\newcommand{\ketbra}[2]{|#1\rangle\langle#2|}
\newcommand{\xbraket}[3]{\langle#1|#2|#3\rangle}
\newcommand{\scalar}[2]{\left\langle{#1}\middle|{#2}\right\rangle}
\newcommand*{\mean}[1]{\mathinner{\langle{#1}\rangle}}

\def\be{\begin{equation}} \def\ee{\end{equation}} \def\bea{\begin{eqnarray}}
\def\eea{\end{eqnarray}}

\newcommand{\dng}{\delta n_g}
\newcommand{\sgn}{\mbox{sgn}}
\newcommand*{\JM}{J_{\rm max}}
\newcommand*{\Jm}{J_{\rm min}}
\newcommand*{\Berry}{\Theta_B}
\def\Jl{J_l}
\def\Jr{J_r}

\begin{document}

\title{Coherent Cooper-pair pumping by magnetic flux control}
% \author{S.~Gasparinetti$^1$ and I.~Kamleitner$^2$}
% \affiliation{${}^1$Low Temperature Laboratory, Aalto University, P.O. Box 15100, FI-00076 Aalto, Finland}
% \affiliation{${}^2$Institut f\"ur Theory der Kondensierten Materie, Karlsruher Institut f\"ur Technologie, 76128 Karlsruhe, Germany}
\author{S.~Gasparinetti}
\affiliation{Low Temperature Laboratory, Aalto University, P.O. Box 15100, FI-00076 Aalto, Finland}
\author{I.~Kamleitner}
\affiliation{Institut f\"ur Theory der Kondensierten Materie, Karlsruher Institut f\"ur Technologie, 76128 Karlsruhe, Germany}

\begin{abstract}
  We introduce and discuss a scheme for Cooper-pair pumping. The scheme relies
  on the coherent transfer of a superposition of charge states across a
  superconducting island and is realized by adiabatic manipulation of
  magnetic fluxes. Differently from previous implementations, it does not
  require any modulation of electrostatic potentials. We find a peculiar
  dependence of the pumped charge on the superconducting phase bias across the pump
  and that an arbitrarily large amount of charge can be pumped in a single cycle
  when the phase bias is $\pi$. We explain these features and their relation to
  the adiabatic theorem.
%   Without a phase bias, the pumped charge approaches half a Cooper-pair per
%   cycle. When the phase bias approaches $\pi$, we find that a single pumping sequence can transfer an
%   arbitrarily large amount of charge. On the other hand, the sequence must be
%   performed correspondingly slower in order to stay in the adiabatic limit. 
\end{abstract}
\pacs{%03.65.Yz, 85.25.Cp, 03.65.Vf
} \maketitle

\section{Introduction}\label{sec:intro}

A Cooper-pair pump \cite{Pekola1999} is a superconducting device that can be
used to transport Cooper pairs by manipulating some of its parameters in a
periodic fashion.
Cooper-pair pumps have recently attracted considerable theoretical
\cite{Pekola1999,Gorelik2001,Aunola2003,Niskanen2003,Fazio2003,Romito2003,Mottonen2006,Cholascinski2007,Safaei2008,Leone2008a,Brosco2008,Pirkkalainen2010,Gasparinetti2011a,Kamleitner2011,Salmilehto2011,Salmilehto2012}
and experimental
\cite{Geerligs1991,Niskanen2005,Vartiainen2007,Mottonen2008,Hoehne2012,Gasparinetti2012}
attention.

Part of this attention stems from the geometric properties of the
parametric cycle used to perform pumping. These
properties, in turn, leave a distinctive fingerprint in the pumped charge. The
link between geometric phases and pumped charge has been established in the
adiabatic limit, where an explicit relation connects the pumped charge to the
Berry phase \cite{Berry1984,Simon1983}, as first shown in
Ref.~\onlinecite{Aunola2003} and experimentally demonstrated in
Ref.~\onlinecite{Mottonen2008}.
In addition, the breakdown of adiabatic behavior due to Landau-Zener
transitions can be detected as a decrease in the pumped charge
\cite{Gasparinetti2012}.
This offers the opportunity to develop Landau-Zener-St\"uckelberg interferometry
\cite{Shevchenko2010}
based on geometric phases \cite{Gasparinetti2011a}.
Finally, it has been proposed to exploit Cooper-pair pumps for the
observation of nonabelian geometric phases \cite{Brosco2008,Pirkkalainen2010}.

Another reason to study Cooper-pair pumps is that they are convenient
solid-state implementations of a driven quantum two-level system. In the
presence of a dissipative environment, the pumped charge is determined by the
quasistationary state reached by the system and thus is a
sensitive probe of decoherence effects. This explains why a Cooper-pair pump was chosen
as a ``case in point'' in several theoretical works aimed at studying the role
of dissipation in driven quantum systems
\cite{Pekola2010,Solinas2010,Russomanno2011,Kamleitner2011}.

Different types of Cooper-pair pumps have been proposed and realized
\cite{Geerligs1991,Niskanen2005,Hoehne2012}. All these devices comprise
the same building blocks, namely, superconducting islands connected to
each other and to superconducting leads by Josephson junctions. They are intended to be
operated in a regime where the charging energy of the islands
is much larger than the Josephson energies of the couplings. Thus, at the heart
of these implementations is the ``classical'' phenomenon of Coulomb blockade.
Pumping relies on periodic modulation of electrostatic potentials, tuned by gate
electrodes which act as pistons in pulling Cooper pairs onto and off the
islands.
The main contribution to pumping comes from incoherent tunneling of Cooper pairs
\cite{Fazio2003}, with phase-coherent effects only providing small corrections.

In this \paper, we undertake a different approach to Cooper-pair pumping, that
we call ``flux pumping'' (FP).
% It is based on the adiabatic, coherent transfer of a superposition of charge
% states across a superconducting island.
% Pumping is achieved by magnetic flux control, while no modulation of
% electrostatic potentials is required.
% While in previous pumping protocols Cooper-pair tunneling is made
% energetically favorable by actiong on the electrostatic potentials, our
% protocol differs in that incoherent tunneling of Cooper pairs. Instead, it
% relies on the creation and maintenance of a superpositions of charge states on
% the island, which can only be created by coherent tunnel processes.�
FP is based on the coherent transfer of a superposition of charge states across
a superconducting island by adiabatic manipulation of magnetic
fluxes.
Contrary to previous proposals, FP does not involve the modulation of
electrostatic potentials.
The pumped charge resulting from FP is purely coherent. Its
dependence on the phase bias across the pump reveals intriguing features. Among
them, we find that for a particular choice of the system parameters, an
arbitrarily large charge can be pumped per cycle. However, this is by no means
inconsistent as at the same time the adiabatic criterion requires the pumping
cycle to be correspondingly slow.

The device we consider for FP uses the same hardware as the Cooper pair sluice
introduced in Ref.~\onlinecite{Niskanen2003}. Yet pumping is achieved in a
completely different manner.
First, the gate voltage is kept constant throughout the pumping cycle.
Second, the opening times of the superconducting quantum interference
devices (SQUIDs) used as valves have a large overlap.
FP also differs from Cooper-pair shuttling
\cite{Gorelik2001,Romito2003}, in at least two respects.
First, while the ``shuttle'' is only coupled to one lead at a
time (hence its name), in our case simultaneous coupling of the central island to both leads is
required to achieve a nonvanishing pumped charge. Second, the operation of the
shuttle is non-adiabatic and requires accurate control of the time dependence of the pulses.
By contrast, FP is insensitive to
the speed at which the cycle is performed, as long as the modulation is adiabatic.
This is a consequence of the geometric nature of the pumped charge.

Features such as the large overlap between the flux pulses, the subordinate role
played by Coulomb blockade, and the overall coherence of the pumping process,
bring FP close together with pumping in open systems,
\cite{Brouwer1998,Switkes1999,Giazotto2011} sometimes referred to as ``quantum
pumping''. FP thus opens a new possibility to explore quantum pumping in
superconducting systems.

% Down a different line, the fact that the central island is essentially a
% superconducting quantum bit makes FP interesting from the point of view of
% quantum information.
% In particular, FP could be used to realize adiabatic transfer of quantum
% states \cite{Vitanov2001}, with the pumped charge providing a convenient proxy
% for the overall fidelity of the process.
% In this spirit, FP may be seen as an implementation of the coherent tunneling
% by adiabatic passage (CTAP) protocol \cite{Eckert2004, Greentree2004,
% Siewert2006, Kamleitner2008}.
% The modulation of the couplings between different charge configurations is
% indeed similar for the two schemes.
% However, the presence of superconducting reservoirs held at a fixed phase
% difference deeply influences FP, leading to peculiar patterns in the
% transferred charge.
% The two schemes also exhibit different symmetries under, e.~g., time reversal.

FP can also be connected to some adiabatic transfer schemes used in quantum
information, in particular, the coherent transfer by adiabatic passage (CTAP)
protocol \cite{Eckert2004, Greentree2004,  Siewert2006, Kamleitner2008}.
In both schemes, the transfer relies on time-dependent manipulation of
tunnelling rates rather than energy levels.
However, only in FP is the device connected to leads, thereby allowing for the
generation of a continuous pumped current. The presence of superconducting leads
and their phase bias introduce novel features that have no equivalent in CTAP.
A hallmark of CTAP is the so-called counter-intuitive pulse ordering: in order
to transfer information from left to right, the right tunnel junction is
operated first.
In FP we do find a similar behavior when the phase bias is close to $\pi$; in
general, however, current flows in the same direction as the pulse sequence.

The outline of this paper is as follows. In Sec.~\ref{sec:sluice}, we introduce
the Cooper-pair sluice and set up the theoretical framework on which our
calculations are based. In the core Sec.~\ref{sec:pumping}, we describe flux
pumping. % We first put forward the more intuitive case where there is no
% superconducting phase bias across the device, and then turn to the general case.
In Sec.~\ref{sec:nonadi}, we characterize the breakdown of adiabatic behavior
by performing numerical simulations with a master equation approach.
Finally, in Sec.~\ref{sec:concl}, we summarize our results and comment on the
feasibility of our proposal.
%In the Appendix, we discuss the relationship between flux pumping and Berry
% phase.

\section{The Cooper-pair sluice}\label{sec:sluice}

\begin{figure}[t]
	\includegraphics[width=\linewidth]{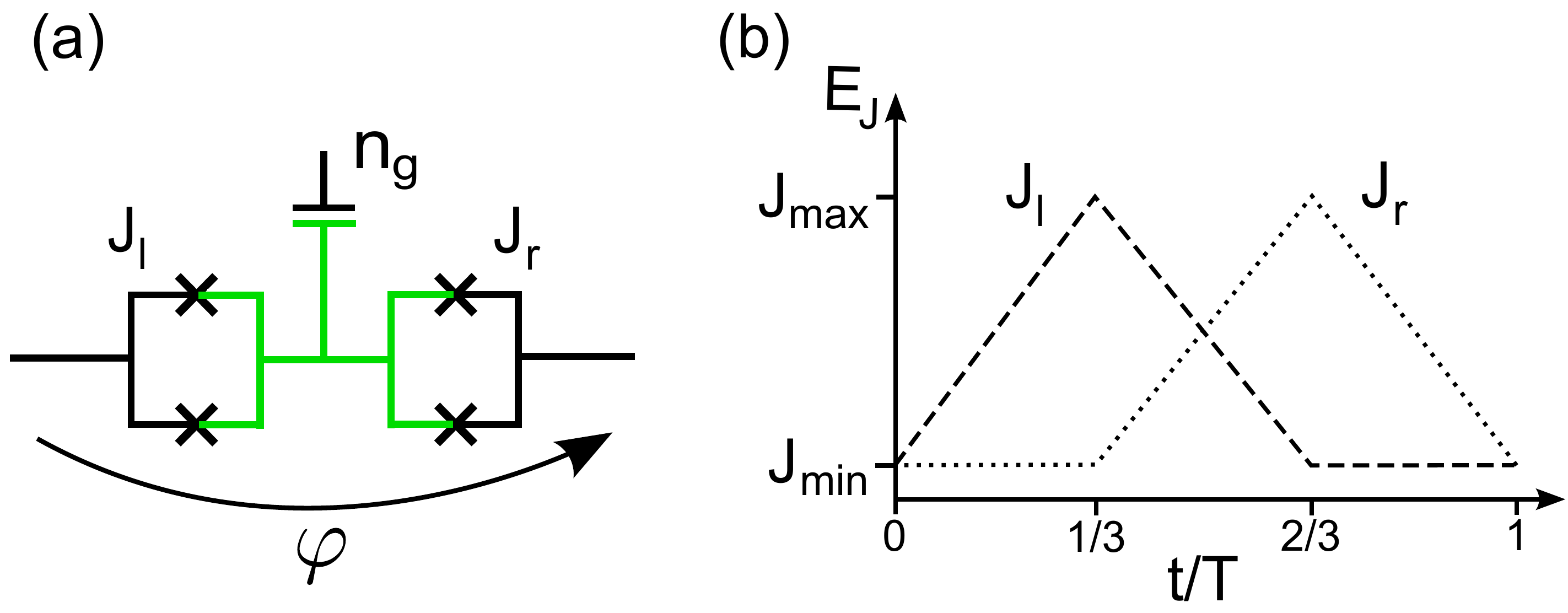} \vspace{-4mm}
	\caption{(Color online) Pumping with magnetic fluxes. (a) Schematic circuit of
	a Cooper pair sluice. A superconducting island (green) is coupled to
	superconducting leads by two SQUIDs, acting as tunable Josephson junctions of
	energy $J_l$, $J_r$. A gate capacitively coupled to the island controls its polarization charge $n_g$.
	The superconducting phase of the two leads is held at a fixed difference $\varphi$.
	(b) Representative time modulation of $J_l$, $J_r$ leading to FP.
	The gate position is kept fixed throughout the modulation.}
	\label{fig:sluice}
\end{figure}

A schematic drawing of the Cooper-pair sluice is shown in
\figref{fig:sluice}~(a). It is a fully tunable Cooper-pair transistor,
consisting of a small superconducting island connected to leads by two
SQUIDs. The SQUIDs are
controlled independently by adjusting the magnetic fluxes $\Phi_l,\Phi_r$
threading their loops, so that they can serve as Josephson junctions of tunable
energy $\Jl$, $\Jr$. A gate electrode capacitively coupled to the island
controls its polarization charge in units of Cooper pairs $n_g=C_g V_g /2e$,
where $C_g$ is the cross-capacitance between gate and island and $V_g$ the gate
voltage.

We assume that the superconducting electrodes to which the sluice is connected
are held at a fixed phase difference $\varphi$.
The simplest way to accomplish this is to embed the sluice in a superconducting loop.
An other possibility is to shunt the sluice with a large Josephson junction of
energy $J_S \gg J_l,J_r$.
In this configuration, the Josephson junction can also serve as a current
threshold detector.
This technique was first applied to the readout of the ``quantronium'' circuit
\cite{Vion2002} and then proposed for \cite{Mottonen2006} and successfully
applied to \cite{Mottonen2008} the sluice.

We use the sluice in the regime where the charging energy $E_C=4e^2/2C_\Sigma$
($C_\Sigma$ is the total island capacitance) is large compared to $\Jl$ and
$\Jr$. We describe the dynamics in the basis of eigenstates of charge on the
island, and restrict the Hilbert space to the states $\ket0$ and $\ket1$ with no
and one excess Cooper pair on the island, respectively. In this two-level
approximation, the Hamiltonian is given in matrix form by \cite{Pekola1999}
% \bea H &\!=\!& \left(\begin{array}{cc} E_C(\frac12+\dng)^2 & \frac12
% \left(\Jle^{-i\varphi/2} + \Jre^{i\varphi/2}\right) \\
% \frac12 \left(\Jle^{i\varphi/2} + \Jre^{-i\varphi/2}\right) &
% E_C(\frac12-\dng)^2 \end{array}\right)\! ,\nn\\
% \eea
\bea
  \hat H &\!=\!& \left(\begin{array}{cc}
    E_C(\frac12+\dng)^2 & J_+ \cos\frac\varphi2 +i J_-\sin\frac\varphi2 \\
    J_+ \cos\frac\varphi2 -i J_-\sin\frac\varphi2 & E_C(\frac12-\dng)^2
    \end{array}\right)\ \ \ \label{eq:Ham}
\eea where $J_\pm = \frac12 \left(\Jl \pm \Jr\right)$, and $\dng=n_g-\frac12$
the offset between the gate charge and the degeneracy point.
% \bea H &\!=\!& -\frac12 \left(\begin{array}{cc} E_C(1-2n_g) &
% \Jle^{-i\varphi/2} + \Jre^{i\varphi/2} \\
% \Jle^{i\varphi/2} + \Jre^{-i\varphi/2} & -E_C(1-2n_g),
% \end{array}\right)\!,\nn\\
% \eea where $n_g=2e V_g/C_g$ and $E_C=2e^2/C_\Sigma$, and $C_g$ and
% $C_\Sigma=C_L+C_R+C_g$ are the capacitances of the gate and the island,
% respectively.

We now outline how to obtain the pumped charge in the adiabatic limit. 
We use the same notation as in Ref.~\onlinecite{Solinas2010}, to which the
reader is referred for a more detailed account.
We also set $\hbar=1$ and $2e=1$.

A pumping cycle is described by a closed loop in the space of a minimal set of
parameters determining $\hat H$. Under the assumption that the parameters are
changed slow enough, the system will approximately follow the instantaneous
ground state of $\hat H$. This fact underlies the adiabatic theorem and is at
the basis of a perturbation expansion.
The latter is formally accomplished by introducing a local adiabatic parameter
\begin{equation}\label{eq:alpha}
\alpha(t) = \left|\braket{\dot{g}(t)}{e(t)}\right| / \Delta(t)\ ,
\end{equation}
where $\ket{g(t)}$ and $\ket{e(t)}$ are, respectively, the instantaneous ground and excited state of
$\hat H$
(adiabatic states)
and $\Delta(t)$ is the instantaneous energy gap at time $t$.
% \begin{equation}\label{eq:alpha}
% \alpha(t)=\lVert D^\dagger (t)\dot D(t)\rVert /\Delta(t)\ , 
% \end{equation}
% where $D(t)$ is the
% matrix that diagonalizes $\hat H$ with respect to a given fixed basis, $\lVert \cdot
% \rVert$ denotes the trace norm and $\Delta(t)$ is the instantaneous
% energy gap in the system at time $t$.
The adiabatic limit is attained provided $\alpha(t) \ll 1$ at all times.

We will find use for the following quantities:
\begin{subequations}\label{eq:quantities}
\begin{align}
E_{12}&=\frac12\sqrt{\Jl^2+\Jr^2+2 \Jl\Jr \cos\varphi}\ , \label{eq:E12} \\
 \gamma &= \arctan\left(  
 \frac{\Jr-\Jl}{\Jr+\Jl}\tan\frac\varphi2 \right)\ , \\
  \eta &=\frac{\dng}{\sqrt{\dng^2+\left(\frac{E_{12}}{E_C}\right)^2}}\ .
 % \omega_0 &= \frac{2E_{12}}{\hbar(1-\eta)}\ .
\end{align}
\end{subequations}
% \bea
% %   b^2 &=& 1-a^2 = \frac12\left[ 1+\frac{\eta}{\sqrt{\eta^2+(E_{12}/E_C)^2}}
% %   \right] \\
% \eea

% A convenient basis to study the adiabatic dynamics is that of the instantaneous
% eigenstates of $\hat H$ (adiabatic states).
In terms of the fixed $\{\ket{0},\ket{1}\}$ basis, the adiabatic states are explicitly given by:
\begin{subequations}\label{eq:g_e}
\begin{align}
\ket{g} &= \frac{1}{\sqrt{2}}\left(\sqrt{1-\eta}\ket0+e^{-i
\gamma}\sqrt{1+\eta}\ket1 \right)\ , \\
\ket{e} &= \frac{1}{\sqrt{2}}\left(\sqrt{1+\eta
}\ket0-e^{-i \gamma}\sqrt{1-\eta}\ket1 \right)\ .
\end{align}
\end{subequations}
% \bea
% %   \ket g = a\ket0 + e^{-i\gamma}b\ket1,\quad \ket e = e^{i\gamma}b\ket0 -
% %   a\ket1,
% \eea
In order to properly account for the pumped charge, it is essential to consider
the corrections to the instantaneous ground state up to first order in $\alpha$.
The resulting density matrix of the sluice, expressed in the adiabatic basis, is given by  
\begin{subequations}\label{eq:rhogg_rhoge}
\begin{align}
\rho_{gg} &= 1\ , \\
\rho_{ge} &= \frac{i\partial _t\eta-\left(1-\eta ^2\right)\partial _t\gamma }{4
E_{12}}\ .
% \rho_{gg} &= 1-e^{-\hbar \omega_0/k_B T} \\
% \rho_{ge} &= \frac{i\partial _t\eta-\left(1-\eta ^2\right)\partial _t\gamma }{4
% E_{12}}(1-2e^{-\hbar \omega_0/k_B T})\ .
\end{align}
\end{subequations}

After solving the dynamics, we can turn to the calculation of the pumped charge.
We introduce current operators $\hat I_k$ for the $k$-th SQUID ($k=l,r$).
% The current operator for
% the $k$-th SQUID (we set $2e=1$ henceforth) is formally given by
% $\hat I_k=\frac{i}{h} \partial_{\varphi_k} \hat H$, where $\varphi_k$ is
% the phase difference across the $k$-th SQUID.
% $\hat I_k=\frac{i}{h}[\hat n_k,\hat H]$, where $n_k$ is the charge passed
% through the $k$-th SQUID.
In the $\{\ket{0},\ket{1}\}$ basis, one has:
\bea \hat I_l &\!=\!& \frac{J_l}{2i} \left(\begin{array}{cc}
    0 & -e^{-i\varphi/2} \\
    e^{i\varphi/2} & 0 \end{array}\right)\ \label{eq:Il} \\
\hat I_r &\!=\!& \frac{J_r}{2i} \left(\begin{array}{cc}
    0 & e^{i\varphi/2} \\
    -e^{-i\varphi/2} & 0 \end{array}\right)\ \label{eq:Ir}
\eea

% Explicitly, one finds:
% \begin{eqnarray} \hat{I}_l=\frac{J_l}{2i}\left( \begin{array}{cc} 0 & e^{i
% \varphi /2} \\
% -e^{-i \varphi /2} & 0 \end{array} \right)\\
% \hat{I}_r=\frac{J_r}{2i}\left( \begin{array}{cc} 0 & -e^{-i \varphi /2} \\
% e^{i \varphi /2} & 0 \end{array} \right) \end{eqnarray} One can also define a
% current operator across the sluice as \bea I &=& 2e\frac{\partial
% H}{\partial\varphi} \nn\\
% &=& i\frac e2 \left(\begin{array}{cc} 0 & \Jle^{-i\varphi/2} -
% \Jre^{i\varphi/2} \\
% -\Jle^{i\varphi/2} + \Jre^{-i\varphi/2} & 0 \end{array}\right) \nn\\
% \eea

The expectation value of the current is given by
$I_k \equiv \tr (\hat \rho \hat I_k)= I_{d,k}+I_{p,k}\ ,$ where we have singled
out a dynamic contribution $I_{d,k}=\rho_{gg}\xbraket{g}{\hat I_k}{g}$ and a
geometric contribution $I_{p,k}=2\Real (\rho_{ge} \xbraket{e}{\hat I_k}{g})$.
% \be I_k = \tr (\hat \rho \hat I_k)= \rho_{gg}I_{k,gg} + \rho_{ee}I_{k,ee} +
% 2\Real (\rho_{ge} I_{k,eg})\ , \label{eq:Ik} \ee where
% $I_{k,\alpha\beta}=\xbraket{\alpha}{I_k}{\beta}$.
While $I_{d,k}$ relates to the usual supercurrent flowing in the presence of a phase bias,
$I_{p,k}$ encapsulates the effects of the parametric drive (note that
$\rho_{ge}=0$ for time-independent parameters) and is thus identified with the
pumped current.
% Explicitly, one finds:
% \begin{subequations}\label{eq:Ipl_Ipr} \begin{align} I_{p,l} &= J_l \left[
% \eta
% \Real(\rho_{ge})\sin(\gamma+\varphi/2)-\Im(\rho_{ge})\cos(\gamma+\varphi/2)\right]
% \nn \\ \label{eq:Ipl} \\
% I_{p,r} &= J_r \left[-\eta
% \Real(\rho_{ge})\sin(\gamma-\varphi/2)+\Im(\rho_{ge})\cos(\gamma-\varphi/2)\right]
% \nn \\ \label{eq:Ipr} \end{align} \end{subequations}
The total charge transferred through the $k$-th SQUID in a period is given by
$Q_{{\rm tr},k}=\int_0^T I_k(t)dt$.
Once again it is possible to distinguish a dynamic charge $Q_{d,k}=\int_0^T
I_{d,k}(t)dt$ and a geometric (pumped) charge $Q_{p,k}=\int_0^T I_{p,k}(t)dt$,
so that $Q_{\rm tr, k}=Q_{d,k}+Q_{p,k}$.
Adiabatic evolution and charge conservation force all three types of charges to
be equal for the left and right SQUID. For this reason, we will safely drop the
subscript $k$ in the following.
% $Q_{{\rm tr},l}=Q_{{\rm tr},r}\equiv Q_{\rm tr}$.

As first shown in Ref.~\onlinecite{Aunola2003}, $Q_d$ and $Q_p$ are related to
the derivative with respect to $\varphi$ of the dynamic phase $\Theta_{\text{d}}$
and the geometric (Berry) phase $\Berry$ accumulated by the instantaneous
ground state along a pumping cycle:
\bea
  Q_d &=& \frac{\partial \Theta_d}{\partial\varphi}\ , \\ 
  Q_p &=& \frac{\partial \Berry}{\partial\varphi}\ . \label{eq:Berry_Qp}
\eea

% The expectation value of the transferred charge per cycle can be calculated by
% \bea Q &=& \int_0^T dt\, \bra{\psi(t)}I(t)\ket{\psi(t)} \\
% &=& \int_0^T dt\, \bra{g}I\ket{g} |\!\scalar{\psi}{g}\!|^2 \nn\\
% &+& \int_0^T dt\, 2\rm{Re}\left[\bra{e}I\ket{g}
% \scalar{\psi}{e}\!\scalar{g}{\psi}\right] \nn\\
% &+& \int_0^T dt\, \bra{e}I\ket{e} |\!\scalar{\psi}{e}\!|^2. \label{eq4} \eea
% The first term in \eqref{eq4} is referred to as the super current $Q_S$ which
% persists even for a time independent Hamiltonian. The second term is the
% pumped charge $Q_p$ which is only non-zero if the excited state $\ket{e(t)}$
% is populated. This population in turn is due to first order adiabatic
% corrections and therefore the pumped current is directly related to the
% time-dependence of the Hamiltonian. The last term is only second order in the
% adiabatic parameter and will be neglected.
Experimentally, $Q_d$ and $Q_p$ can be distinguished as they obey different
symmetries. In particular upon reversing the direction of the pumping, $Q_d$ is
not affected, while $Q_p$ duly changes its sign. We shall henceforth
assume that such a distinction can be made and only be concerned with $Q_p$.

% We will use \eqref{eq6} to calculate the pumped charge quantitatively, but we
% will also use \eqref{eq4} for a qualitative, but more intuitive discussion.

\section{Flux pumping}\label{sec:pumping}

The principle of FP is illustrated in Fig.~\ref{fig:sluice}~(b), showing the
time evolution of the control parameters during a pumping cycle.
The charge offset $\dng$, not shown, is set close to the degeneracy point
(that is, $|\dng| \ll 1$)
and kept fixed throughout the cycle. At the initial time $t=0$, the island is
decoupled from both leads, with $\Jl$ and $\Jr$ set to their minimum value
$\Jm$. In sector I ($0 \leq t < T/3$), the coupling to the left lead is turned
on by maximizing $\Jl$. In sector II ($T/3 \leq t < 2T/3$), the coupling is
swapped from the left to the right lead, in such a way that the sum $J_l+J_r$ is
kept constant.
Finally, in sector III ($2T/3 \leq t < T$) $\Jr$ is turned down to $\Jm$,
bringing the system back to the initial state.
We have chosen linear ramps and a perfect coupling swap only for simplicity
in deriving analytical expressions. As it goes with geometric pumping, a
moderate tweaking of the pulses will not disrupt the pumping process as long as
the solid angle spanned in the parameter space [see App.~\ref{app:Berry}] stays
approximately the same.

In the following, we will use the formalism of Sec.~\ref{sec:sluice} to
understand the adiabatic dynamics generated by FP and the corresponding pumped
charge. We first discuss the case $\varphi=0$, as it allows for an intuitive
explanation. For simplicity, we also set $\Jm=0$.

\begin{figure}[t]
	\includegraphics[width=\linewidth]{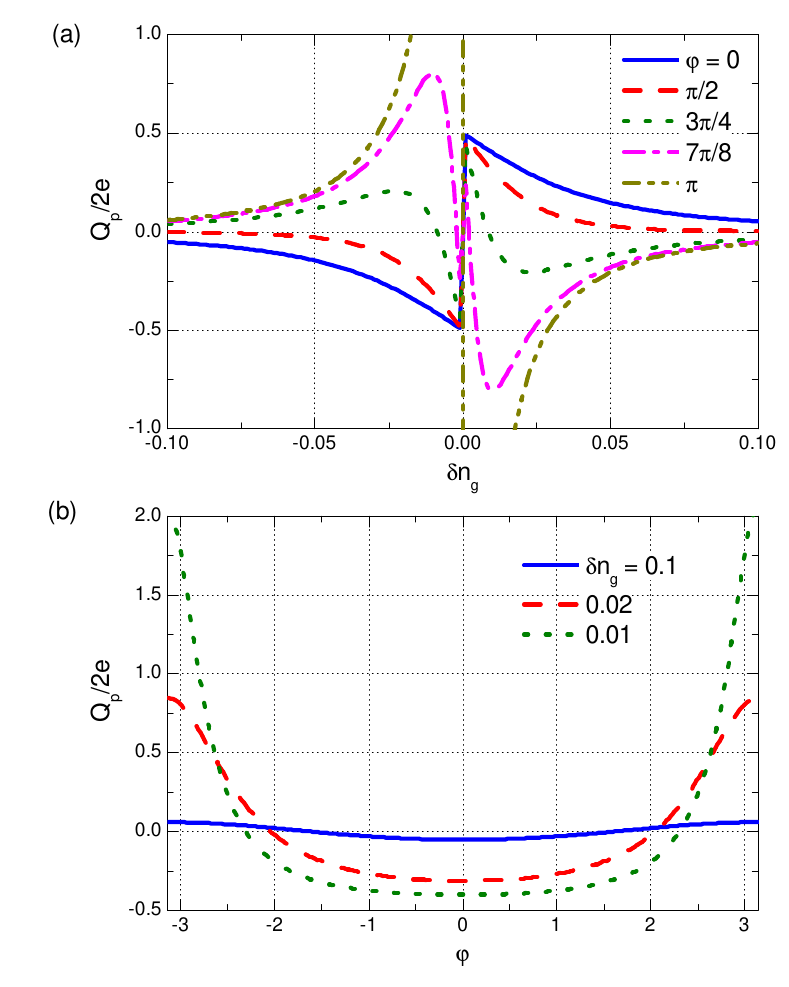} %\vspace{-4mm}
	\caption{(Color online)
	Pumped charge in the adiabatic limit.
	(a) Pumped charge $Q_p$ versus offset charge $\dng$ for different values of the
	superconducting phase bias $\varphi$.
    (b) $Q_p$ versus $\varphi$ for different values of $\dng$.}
    \label{fig:adi}
\end{figure}

\subsection{The case $\varphi=0$}\label{sec:pumping:phi=0}

At $t=0$, the island is in a definite charge state (0 if $\dng<0$, 1 if
$\dng>0$) and the energy difference between charge states is $2E_C\dng$. As
$\Jl$ increases so that $\Jl \gtrsim E_C\dng$, the ground state evolves into a
superposition of charge states.
As a result, charge flows from the left lead onto the island. The charge
transferred in this case is simply given by $|\braket{1}{g(\frac T3)}|^2-
|\braket{1}{g(0)}|^2$.
% \add{ Notice that the maintenance of coherence between different charge states
% is essential in order to achieve a net charge transfer; if tunnelling of
% Cooper pairs were incoherent, no charge would be transferred. See discussion
% in App.~\ref{app:extr_deph}.}
In sector II, the swapping of the couplings does not change the Hamiltonian
\eqref{eq:Ham}. As a result, $\rho_{ge}=0$ and no charge is transferred.
Finally, in sector III, the same amount of charge is released to the right lead
as the system comes back (up to a geometric phase) to the initial state.
Some plots of the instantaneous geometric currents $I_{p,l},I_{p,r}$ are shown
in App.~\ref{app:curr_plots}.

In this scheme, the fact that the dynamics is coherent plays a crucial role.
This marks a clear difference between FP and previous pumping protocols. In the latter
Cooper-pair tunneling is made energetically favorable via modulation of
electrostatic potentials, so that the coherent-versus-incoherent nature of
the tunneling process has only a modest influence on the pumped charge.
By contrast, in FP there are no ``pistons" pulling the Cooper-pairs around.
As a result, one can show that in the limit of incoherent Cooper-pair
tunneling the pumped charge vanishes.

The total pumped charge for the case $\varphi=0$ can be inferred from this
heuristic argument, calculated by direct integration of
$I_{p,k}$ [see Sec.~\ref{sec:sluice}], or obtained by virtue of
\eqref{eq:Berry_Qp} [see App.~\ref{app:Berry}]. The result is:
\begin{equation}
Q_{p}[\varphi=0]  = - \frac12 \sgn(\dng)
\left(1-\frac{1}{\sqrt{1+r^2}}\right)\ , \label{eq:Qp_I_phi0}
\end{equation}
% \frac{r^2}{1+r^2+\sqrt{1+r^2}}\ ,
where we have introduced the ratio
\begin{equation}\label{eq:r}
r=\frac{\JM}{2E_C\dng}\ .
\end{equation}
In the limit $r \to \pm \infty$ (corresponding to $E_C|\dng| \ll \JM$), the
absolute value of $Q_p$ approaches a maximum of half a Cooper pair. This result is
approximately valid also for a finite $\Jm$, as long as $\Jm \ll E_C\dng$.

The dependence of $Q_p$ on $\dng$ exhibits a sawtooth behavior, as shown by
the solid line in \figref{fig:adi}(a). At $\dng=0$, the pumped
charge changes sign discontinuously. However, when $\dng \to 0$ also the minimum energy gap
$\Delta E_\text{min} \equiv \min_{t \in [0,T]} \Delta E(t) = \dng E_C$ tends to
zero. This implies that the adiabatic limit, in which the present derivation
is valid, is only attained for infinitely slow evolution. We will return to this point in
Sec.~\ref{sec:nonadi}.

% For $\varphi=0$ the supercurrent vanishes.
% and the simple picture explained above with moving charges on and off the island
% applies.
% For $\varphi\neq 0$, there is not only an additional charge $Q_S$ due to the
% super current, but also the pumped charge $Q_p$ can not be explained by these
% means as we will see below.

\subsection{The general case}\label{sec:pumping:gen}

When $\varphi\neq 0$, the same calculation leading to \eqref{eq:Qp_I_phi0} shows
that the pumped charge in sectors I and III is the same as in the
case $\varphi=0$. This is only to be expected: as long as the island is
only coupled to a single lead, the phase difference between the leads
cannot play any role.
The situation is different for sector II, where the coupling swap now takes
place between two leads at different phases. As a result, an adjustment of the
superconducting order parameter on the island is required. \cite{phase_note}
This causes an additional geometric current to flow across the
sluice, in a direction opposite to that of the pumping.

In \figref{fig:adi}(a) we plot $Q_p$ versus $\dng$ for different values of
$\varphi$. For values of $\varphi$ in the range of $0$ and $\pi/2$, $Q_p$ simply
decreases with respect to the case $\varphi=0$. As $\varphi$ is further increased, however, a new trend
emerges: $Q_p$ changes its sign with respect to the unbiased case, except in the
vicinity of the degeneracy point. The magnitude of the counterflowing $Q_p$ can
well exceed a Cooper pair. Finally, at $\varphi=\pi$ the sign of the pumped
charge is opposite to that of the unbiased case for all values of $\dng$.
Furthermore, $Q_p$ diverges as $1/\dng$ for $\dng \to 0$.

The full dependence of $Q_p$ on $\varphi$ is shown in \figref{fig:adi}(b) for
three selected values of $\dng$.
The reader may guess that the integral of each curve in \figref{fig:adi}(a)
vanishes. Indeed, using \eqref{eq:Berry_Qp} we obtain $\mean{Q_p}_\varphi \equiv
\frac{1}{2\pi}\int_0^{2\pi} Q_p(\varphi) d\varphi= \Berry(2\pi)-\Berry(0)$.
Even if in general $\Berry$ does not have to be single-valued, \cite{Mottonen2006}
in the present case $\Berry(2\pi)=\Berry(0)$, so that
$\mean{Q_p}_\varphi=0$. This implies that FP can only be observed in
the presence of a well-defined phase bias, for if $\varphi$ randomly fluctuates
in time, then no net charge is transferred on average.
This is a clear signature of the coherent nature of the pumping process.
On the other hand, $Q_p$ exhibits some degree
of robustness against small phase fluctuations. In particular, for $\dng \ll 1$,
$Q_p$ develops a plateau centered at $\varphi=0$. This can
be seen in the increased flattening of the curves with smaller $\dng$ in
Fig.~\ref{fig:adi}(b) (an analytical argument is provided in App.~\ref{app:Berry}).

We remark that these features are peculiar to FP.
To draw a comparison, let us recall that in ``ordinary'' Cooper-pair pumping
\cite{Niskanen2003,Mottonen2006} $\mean{Q_p}_\varphi=1$, the phase dependence
of $Q_P$ only appears as first-order correction in the small parameter $\Jm/\JM$, and no significant
dependence on the charge offset $\dng$ is found as long as the gate modulation crosses the degeneracy
point.

\subsection{The case $\varphi=\pi$}\label{sec:pumping:phi=pi}

We now fix our attention on the case $\varphi=\pi$, for which we can
present analytical results. Upon direct integration of the
current operator, we obtain for the charge pumped in the sector
II
\begin{equation}
Q_p^{(\text{II})}[\varphi=\pi]  = \sgn(\dng)\frac{r^2}{2 \sqrt{1+r^2}} \ .
\end{equation}
This must be added to the contribution \eqref{eq:Qp_I_phi0} from sectors
I,III, to give the total pumped charge
\begin{equation} \label{eq:Qp_I_phi_pi}
Q_{p}[\varphi=\pi] = \frac12 \sgn(\dng) \left(\sqrt{1+r^2}-1\right)\ .
\end{equation}
An alternative derivation of \eqref{eq:Qp_I_phi_pi} involving the Berry phase is
shown in App.~\ref{app:Berry}.
On comparing \eqref{eq:Qp_I_phi0} and \eqref{eq:Qp_I_phi_pi}, we see that the
pumping direction for $\varphi=\pi$ is always opposite to that for $\varphi=0$.
From \eqref{eq:Qp_I_phi_pi}, it is apparent that $Q_p$ diverges for
$\varphi=\pi$ and $r\to\infty$ (or $\dng\to0$). Notably, one finds that this
divergence is not removed even when relaxing the constraint $\Jm=0$.
 
As mentioned, the present results have been derived in the adiabatic limit.
So far, we have not investigated how tight a requirement this imposes on the
driving frequency.
To do so, we first notice that the adiabatic condition $\alpha \ll
1$ is equivalent to $\rho_{ge} \ll 1$.
Now for $\varphi=\pi$, $|\rho_{ge}|$ is maximum at $t=T/2$, where it attains the
value \be \label{eq:maxrhoge_phi=pi} \max_{0\le t < T} |\rho_{ge}| = \frac{3
r^2}{\JM T}\ .
\ee
% We thus see that $|\rho_{ge}|$ also diverges for $r \to \infty$.
Eq.~\eqref{eq:maxrhoge_phi=pi} implies that when approaching the degeneracy point, the pumping period
should be increased according to $T \propto r^2$ in order to stay in the
adiabatic limit. In other words, the increase in $Q_p$ comes at the cost of an
increasingly long $T$.
We discuss this point in more detail in App.~\ref{app:optimal}.
% This discussion can be made quantitative: as we show in
% App.~\ref{app:optimal}, Eqs.~\eqref{eq:Qp_I_phi_pi} and \eqref{eq:maxrhoge_phi=pi} correctly
% predict the scaling of the adiabatic breakdown.
One can also check that the
pumping current $I_p$ does not diverge at any time. In fact, $I_p$ never exceeds
$I_{\rm max}\approx \frac{\JM}{4}\rho_{ge} \ll \JM$, that is, much less than
the critical current of the SQUIDs.

% The average pumped current $I_p=Q_p/T$ scales as $I_p \propto r^{-1}$.

% In fact, the period $T$ should be large compared to the
% inverse of the minimal transition frequency
% $\min_{t\in(0,T)}(E_e-E_g)/\hbar=2E_C \eta/\hbar$, which means that near the
% degeneracy point the period has to diverge with $1/\eta$ for adiabatic pumping.
% Therefore, the pumped current is limited by \be I_P \ll
% \frac{E_{\text{max}}e}\hbar. \ee
% 
% The divergence of the pumped charge per cycle can be understood from
% \eqref{eq:rhoge}. As we have seen, the pumped charge is due to the term
% $\scalar{\psi}{e}\!\scalar{g}{\psi}$. The population of the excited state is
% only due to first order adiabatic corrections\cite{note:adiab} which are linear
% in $\frac{\scalar{\dot g}{e}}{(E_e-E_g)/\hbar}$. For $\varphi=\pi$ we have the
% special situation that at $t=T/2$ the energy splitting linearly approaches zero
% as we move towards the degeneracy point, therefore resulting in a large
% population of the excited state, or, better, in a constant population
% if $T$ increases as $1/\eta$. From this point of view it is clear that the
% pumped charge per cycle diverges at the degeneracy point and for $\varphi=0$.
% Even though the maximum speed to keep in order to stay in the adiabatic limit
% goes to zero.

\section{Adiabatic breakdown and decoherence}\label{sec:nonadi}

In Sec.~\ref{sec:pumping}, we carried out our calculations assuming adiabatic
evolution during the pumping cycle. In this limit, the pumped charge is uniquely
determined by the loop described in the parameter space of $\hat H$.
As such, it does not depend on the pumping frequency.
We also pointed out, however, that the validity of the adiabatic theorem
requires the condition $|\rho_{ge}|\ll 1$ to hold. We then warned the reader
that in the limit $\dng \to 0$, due to the vanishing of the instantaneous energy
gap at $t=0$ (and at $t=T/2$ when $\varphi=\pi$), this condition requires the
pumping cycle to be infinitely slow. As real measurements are always performed
at finite frequencies, this implies that none of the traces of
\figref{fig:adi}(a) can be exactly reproduced in an experiment in the vicinity
of $\dng=0$.

In this Section, we investigate the behavior of the pumped charge beyond the
adiabatic limit. As soon as we allow nonadiabatic transitions to take place,
the dynamics of the system becomes highly nontrivial. As argued elsewhere
\cite{Pekola2010,Solinas2010,Russomanno2011,Salmilehto2011,Gasparinetti2011a}, the pumped
charge is a sensitive probe of this dynamics. Indeed, it was exploited in
Ref.~\onlinecite{Gasparinetti2012} to characterize Landau-Zener transitions in
the Cooper-pair sluice.
Furthermore, the final state of the pump at the end of the pumping cycle is in
general different from its initial state. As measurements are typically averaged
over very many cycles, the quantity of experimental interest
becomes
the stationary pumped charge $Q_p^{\rm st}$. The latter is
determined by an interplay between the nonadiabatic drive and decoherence
effects due to the electromagnetic environment in which the pump is embedded.
A full characterization of such effects is beyond the
scope of this work. Still, the inclusion of decoherence in the model is
essential in order to reach a quasistationary state.

\begin{figure}[t] \includegraphics[width=\linewidth]{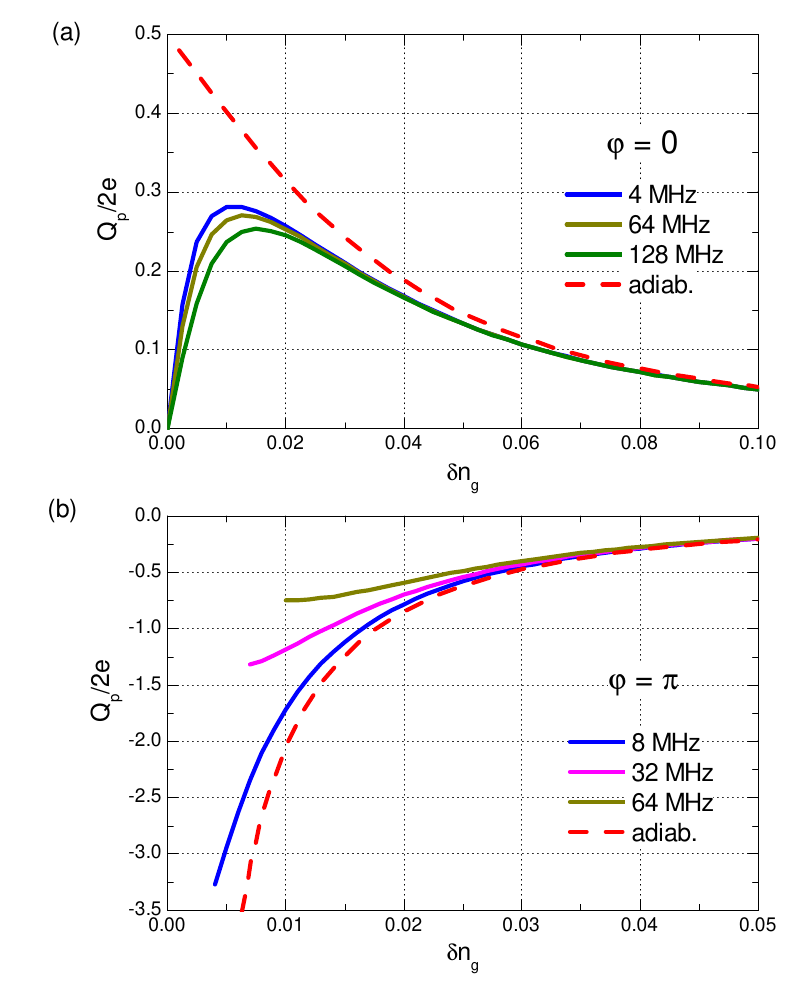} %\vspace{-4mm}
\caption{(Color online) Breakdown of adiabatic behavior.
$Q_p$ versus $\dng$ for $\varphi=0$ (a) and $\varphi=\pi$ (b), for different
pumping frequencies $f$ (solid lines). The results are obtained by numerical
integration of the master equation of Ref.~\onlinecite{Pekola2010}. The
adiabatic-limit predictions \eqref{eq:Qp_I_phi0}
and \eqref{eq:Qp_I_phi_pi} are also plotted for comparison (dashed lines).
Smooth pulses are used in place of those of \figref{fig:sluice}(b) for improved
adiabaticity. Relevant parameter values for the pump are: $E_C=\SI{1}{K},
\JM=0.1 E_C, \Jm=0.03\JM$. For the fictitious environment (see
Ref.~\onlinecite{Solinas2010} for details): $g=0.02$, $R=\SI{300}{k\ohm}$,
$T=0$, $T_0=\SI{0.4}{K}$.}
	\label{fig:nonadi}
\end{figure}

We present numerical results obtained using the master equation approach
developed in Refs.~\onlinecite{Pekola2010,Solinas2010}, which consistently
accounts for the combined action of a quasi-adiabatic drive and decoherence on
an open quantum two-level system. This formalism is not intended to address the
fully nonadiabatic case; yet, it can be conveniently used to investigate the
parameter region where the adiabatic condition ceases to hold.
Decoherence (dephasing and relaxation) is modeled by attaching a fictitious
environment to the pump, in the form of a resistor capacitively
coupled to the central island. This mimics the effect of charge noise, which is
known to be the first cause of decoherence in charge-based devices
\cite{Ithier2005}.

In Refs.~\onlinecite{Pekola2010,Solinas2010} it was found that a
zero-temperature environment tends to stablize ground-state pumping, effectively
extending the region of adiabaticity. Here we also consider a zero-temperature
enviroment. By tuning the coupling parameter to a small value, we make sure that
nonadiabatic transitions still play the major role. In this case, decoherence
only acts as a weak source of dissipation: it damps the oscillations in the
pumped charge and slowly brings the system into a quasistationary state.
% The coupling parameter, determined in the model by a ratio of capacitances, is
% chosen to be very weak, so that nonadiabatic transitions prevail over
% environment-induced relaxation \cite{Pekola2010}.
% % In the following, however, we will focus on the breakdown of the adiabatic %
% limit rather than the influence of the environment.

In \figref{fig:nonadi} we plot $Q_p^{\rm st}$ versus $\dng$ for the emblematic
cases $\varphi=0$ (a) and $\varphi=\pi$ (b).
We choose the realistic device parameters $E_C=\SI{1}{K}, \JM=0.1 E_C,
\Jm=0.03\JM$, use smooth pulses instead of those in \figref{fig:sluice}(b) and
explore different pumping frequencies (solid lines).
The adiabatic-limit predictions for the two cases (Eqs.~\ref{eq:Qp_I_phi0} and
\ref{eq:Qp_I_phi_pi}, respectively) are also plotted for comparison (dashed
lines); notice, however, that they were derived in the limit $\Jm\to0$.

In general, nonadiabatic transitions result in a decrease of $Q_p$. This is
qualitatively accounted for by the fact that the adiabatic excited state of the
sluice carries an opposite $Q_p$ with respect to the ground state.
For the case $\varphi=0$ [\figref{fig:nonadi}(a)], this results in a smearing of
the adiabatic sawtooth. Note that since we have considered the realistic case
$\Jm \neq 0$, in the limit $\dng\to 0$ one still has $\Delta E_\text{min}=\Jm$,
so that the residual coupling partly holds back nonadiabatic transitions.
Analogous considerations can be made for the case $\varphi=\pi$
[\figref{fig:nonadi}(b)]. As $\dng$ is reduced, however, the nonadiabatic
behavior is no longer mitigated by the presence of a finite $\Jm$ (this relates
to the fact that $E_{12}$ vanishes for $J_l=J_r$  when $\varphi=\pi$, see
Eq.~\ref{eq:E12}).
The effect is thus more dramatic, with higher frequencies hitting the
nonadibatic onset first. We terminate each data series as soon as
$\rho_{ge}$ exceeds the arbitrary threshold 0.3; further points would fall
outside the range of validity of our master equation.

These results indicate that nonadiabatic effects must be taken into
serious consideration in any pratical implementation of FP.
A relevant figure of merit for optimization is the average geometric current
$\mean{I_p}=fQ_p$ ($f=1/T$), as this is the signal to be detected in a realistic readout
scheme. An example of such optimization is presented in App.~\ref{app:optimal}
for the case $\varphi=\pi$.

\section{Conclusions}\label{sec:concl}

We have presented a new scheme for Cooper-pair pumping, flux pumping (FP).
Based on magnetic-flux control, FP uses neither a bias voltage nor a modulation
of gate voltages. FP is realized by coherent transer of a
superposition of charge states across a superconducting island.
The resulting pumped charge depends on the gate position and on the phase
difference across the pump in a distinctive fashion.
As no incoherent process can mimic these features, their witnessing
would be an unambiguous demonstration of purely coherent Cooper-pair pumping.
% Furthermore, the similarity between FP and other adiabatic state transfer protocols
% suggests that FP may find applications in quantum information.

The implementation of FP looks feasible, expecially in the light of recent results
obtained with the Cooper-pair sluice \cite{Mottonen2008,Gasparinetti2012}.
An apparent matter of concern is the fact that the supercurrent flowing through
the device may well exceed the pumped current. For instance, let us take the
device parameters of \figref{fig:nonadi} and the pumping cycle of
\figref{fig:sluice}(a). The mean dynamic current at $\varphi=\pi/2$ can be
approximated by $\mean{I_d}\approx\frac{2e}{24\hbar}\JM\approx \SI{350}{pA}$,
independent of frequency. Now at a typical
$f=\SI{120}{MHz}$, $Q_p=e$ corresponds to $\mean{I_p} \approx \SI{20}{pA}$, so
that the pumped current accounts for less than $10\%$ of the signal.
This is not an issue, however, as the supercurrent term is even with respect to
time-reversal symmetry, while the pumped current is odd. As a result, $\mean{I_p}$ can
be determined by simply subtracting the measured currents when pumping
in opposite directions (as done in Ref.~\onlinecite{Mottonen2008}).

Still, detecting such a small current circulating in a loop may challenge
customary techniques. In the search for signatures of FP, an important role is
likely to be played by its distinctive symmetries. Besides time-reversal,
$\mean{I_p}$ is also an odd function of the gate position with respect to
degeneracy. Finally, it should not depend on the direction of the circulating
currents in the SQUIDs. Altogether, these symmetries may be used to rule out the
contribution of undesired rectification effects, possibly originating from
spurious inductive or capacitive couplings.

\acknowledgments

The authors are grateful to T.~Aref, L.~Arrachea, F.~Giazotto, J.~Pekola,
A.~Shnirman, and P.~Solinas for valuable discussions.
This work was supported by the European Community FP7 under grant No.~238345
``GEOMDISS''. S.~G.~acknowledges financial support from the Finnish National
Graduate School in Nanoscience.

\appendix

% \add{
% \section{Flux pumping in the presence of extreme charge dephasing}\label{app:extr_deph}
% In this section we show that the maintaining of a definite phase coherence between different
% charge states is a necessary condition for flux pumping.
% To do so, let us consider a case of extreme charge dephasing by assuming that the
% coherence between different charge states is destroyed at any given instant.
% We model this by applying the following transformation to the density matrix:
% $$
% \rho \to \rho'\equiv \sum_{i=0}^1 P_i \rho P_i\ ,
% $$ 
% where $P_i=\ketbra{i}{i}$.
% We find
% \begin{subequations}
% \begin{align}
% \rho_{gg}' &= 1-\frac12\left[1-\eta^2+2\Realal \left(\rho_{ge} \right)  \eta \sqrt{1-\eta^2} \right]\ , \\
% \rho_{ge}' &= \Real \left( \rho_{ge} \right) (1-\eta^2)-\eta \sqrt{1-\eta^2}\ .
% \end{align}
% \end{subequations}
% We first notice that if $\ket{g}$ is a charge eigenstate, than $\rho'=\rho$. This corresponds to the cases $\eta=\pm 1$.
% We then notice that only the real part of $\rho_{ge}$ survives the projection. This means that the imaginary part of $\rho_{ge}$,
% responsible for the charge transfer in sectors I and III, now plays no role.
% As we expect, no current flows in Sectors I and III in this case. By contrast, current keeps flowing for finite values of $\varphi$ in Sector II.
% This is because this current is associated with a rearrangement of the phase on the island, and as such is more robust against charge dephasing.
% 
% This is too drastic, however: if we apply it the ordinary sluice, we get that no charge is transferred there, either.
% }

\section{Instantaneous geometric currents}\label{app:curr_plots}

\begin{figure}
   \includegraphics{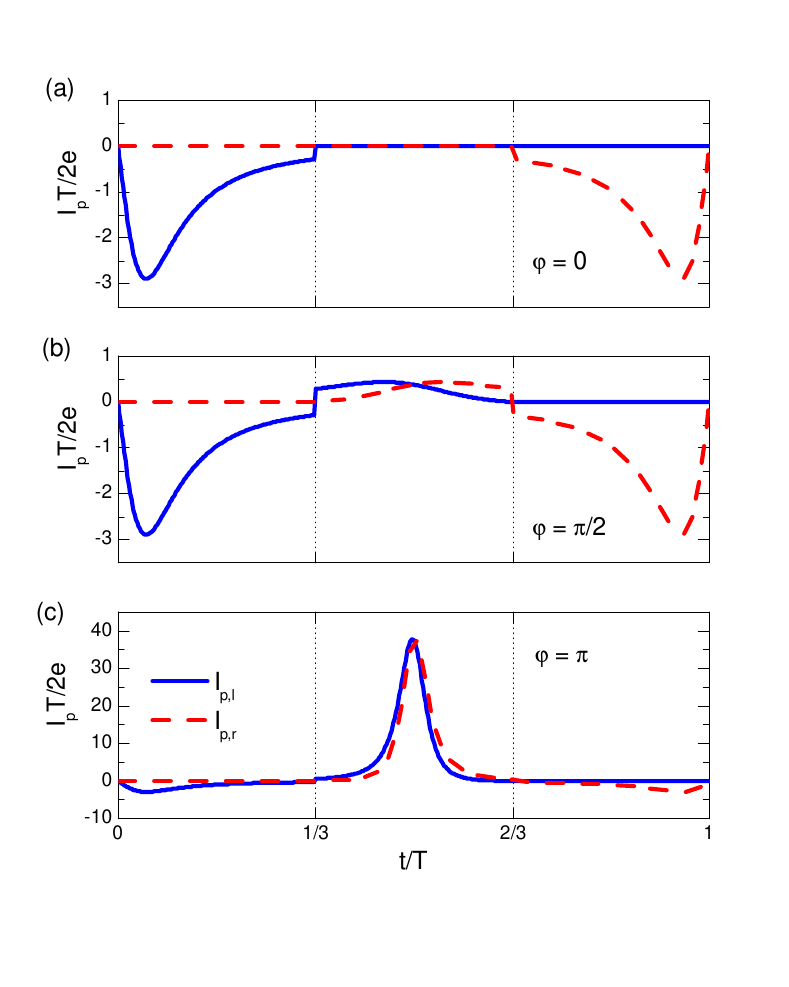}
   \caption{(Color online)
   Instantaneous geometric currents. $I_{p,l}$ (solid line) and $I_{p,r}$ (dashed line) versus time 
   for $\dng=-0.02$ and $\varphi=0$ (a), $\pi/2$ (b), and $\pi$ (c).
%    To avoid unphysical discontinuities in the current, we have used smooth pulses
%    equivalent to those of \figref{fig:sluice}(b).
   }
	\label{fig:currs}
\end{figure}

In \figref{fig:currs} we plot the instantaneous geometric currents $I_{p,l}$ and
$I_{p,r}$ for the cases $\varphi=0$ (a), $\pi/2$ (b), and $\pi$ (c).
The current profiles in the first and third sector are the same for all three plots
[notice the change of scale in panel (c)]. By contrast, the currents in sector II strongly depend
on $\varphi$. A counterflowing current, absent for $\varphi=0$ [panel (a)] develops for finite $\varphi$ [panel (b)]
and largely exceeds the forward current in magnitude as $\varphi$ approaches $\pi$ [panel (c)].
The discontinuities at sector boundaries are due to the cusps in the pulses of
\figref{fig:sluice}(b) and disappear as soon as the latter are replaced by
smooth pulses.

\section{Flux pumping and Berry phase}\label{app:Berry}

\begin{figure}
   \includegraphics[width=0.9\linewidth]{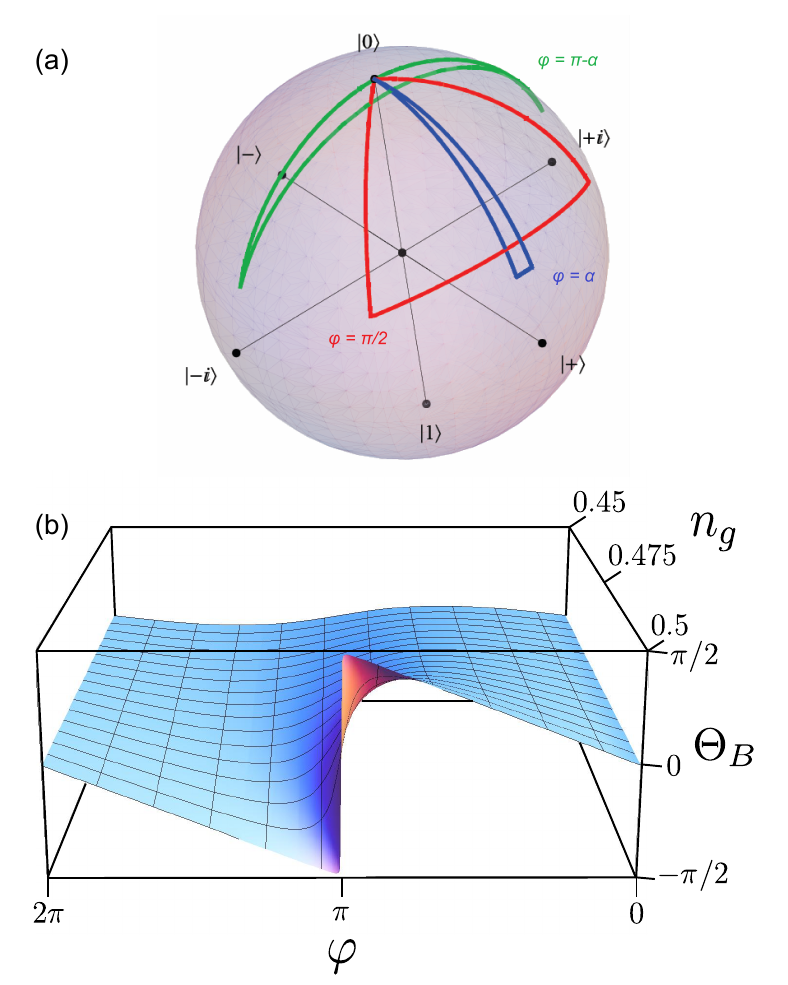}
   \caption{(Color online) Pumping cycle and Berry phase.
   (a) Plots of the
   path drawn by the ground state of the sluice on the Bloch sphere along a
   pumping cycle, for $\varphi=\alpha$ (blue), $\varphi=\pi/2$ (red), and
   $\varphi=\pi-\alpha$ (green), with $\alpha \ll 1$. The Berry phase is
   proportional to the solid angle spanned by the paths.
   (b) Berry phase $\Berry$ versus $\varphi$ and $\dng$. According to
\eqref{eq:Berry_Qp}, The pumped charge $Q_p$ is proportional to the slope of
the surface plot along the $\varphi$ axis.}
	\label{fig:Berry}
\end{figure}

The charge pumped by a
superconducting pump in the adiabatic limit is linked to the Berry phase
$\Theta_B$ accumulated by the instantaneous ground state along the pumping cycle
\cite{Aunola2003}, as prescribed by \eqref{eq:Berry_Qp}.
For a two-level system parametrically driven in closed loop, $\Theta_B$ is
proportional to the solid angle spanned by the Bloch vector, which performs an
adiabatic rotation on the Bloch sphere. The path drawn by the Bloch vector is shown in \figref{fig:Berry}(a) for the pumping cycle of
\figref{fig:sluice}(b) and a few selected values of $\varphi$.
The resulting $\Berry$ is plotted versus $\varphi$ and $\dng$ in
\figref{fig:Berry}(b). According to \eqref{eq:Berry_Qp}, the pumped charge $Q_p$
is given by the local slope of the surface plot along the $\varphi$ axis.

Using \eqref{eq:quantities}, \eqref{eq:g_e}, and the definition
\cite{Berry1984}, we find
\begin{equation} \label{eq:Berry}
  \Berry \equiv i\int_0^T dt \scalar{g}{\dot g} 
  = \frac12 \int_0^T dt \left(1+\eta\right) \dot\gamma\ .
\end{equation}

For the given pumping cycle, $\gamma(t)=-\varphi/2$ in sector I and
$\gamma(t)=\varphi/2$ in sector III. The time derivative of $\gamma$ vanishes in
these regions. The sudden change of $\gamma$ from $\varphi/2$ to $-\varphi/2$ at
times $0,T,\ldots$ yields a delta function in the time derivative, but at
that time $1+\eta=0$, so that the integrand in \eqref{eq:Berry} vanishes as well.
As a result, the only contribution to $\Berry$ comes from sector II. The
fact that sectors I and III do not contribute to $\Berry$ is a consequence of
our choice of adiabatic basis (Eqs.~\ref{eq:g_e}), and is not in contrast with
the fact that there is a charge flow in sectors I, III. Indeed, $\Berry$
is only defined for closed loops. It is possible to give a gauge-invariant
generalization of the Berry phase for open loops \cite{Samuel1988},
but we do not need it here.

We will now explicitly calculate $\Berry$ and $Q_p$ in two
important cases.

% \bea
%   \theta_{\text{geo}} &=& \int_{T/3}^{2T/3} dt\, b^2\dot\gamma %\nn\\
%   % &=& \frac12\int_0^1 du\, \frac{d}{du}\left\{ \arctan\left[
%   % (1-2u)\tan\frac\varphi2 \right] \right\} \nn\\
%   % && \times \left[ 1+\frac{\eta}{\sqrt{\eta^2+ (c/2)^2 [
%   % 1-2(1-\cos\varphi)u(1-u) ]  }} \right] \nn\\
% \eea

\subsection{Case $\varphi=0$}

For definiteness, we assume $\dng<0$.
To calculate $Q_p$ for $\varphi=0$, it is sufficient to expand $\Berry$ to first
order in $\varphi$.
Up to this order, $\eta$ is time-independent in sector II, so that 
$\Berry\approx \frac12 (1+\eta)[\gamma(2T/3)-\gamma(T/3)]=\frac12
(1+\eta)\varphi$. The pumped charge is thus
\be
  Q_p[\varphi=0] = \frac12(1-\eta)\ , \label{eq:app:Qp0}
\ee
as we found in \eqref{eq:Qp_I_phi0}.

It is worth noting that near the degeneracy point $\dng \ll 1$, the validity of
\eqref{eq:app:Qp0} extends to all phases $\varphi \neq \pi$. In fact, for sufficiently
small $\dng$, $\eta \approx 0$ for all times in sector II. As a result, the pumped
charge is half a Cooper pair, as predicted by \eqref{eq:app:Qp0}.

\subsection{Case $\varphi=\pi$}

Using \eqref{eq:Berry_Qp} and \eqref{eq:Berry}, we find for the pumped charge
\bea Q_p[\varphi=\pi] &\!=\!& \frac12 \int_{T/3}^{2T/3}\!\!\! dt \left[
  1+\frac{\dng}{\sqrt{\dng^2+[(\Jl-\Jr)/(2E_C)]^2}} \right] \!\nn\\
  && \times \frac{\partial^2}{\partial t\, \partial\varphi} \arctan\left(
  \frac{\Jr-\Jl}{\Jr+\Jl}\tan\frac\varphi2 \right).
\eea Using $\Jr+\Jl=\JM$ and
$\Jr-\Jl=6\JM u/T$ with $u=t-T/2$, the integral can be
evaluated analytically to give
\begin{equation}
\begin{split}
  Q_p[\varphi=\pi] &\!=\! \frac{T}{12} \int_0^{T/6} \! \frac{du}{u^2} \left[ 1-
  \frac{1}{\sqrt{1+(6 r u/T)^2 }} \right]\!  \\
  &= \frac12 \left[ \sqrt{1+r^2}-1 \right]\ .
%   &\approx \left\{ \begin{array}{cl}
%     r/2 & \mbox{for } r \gg 1 \\
%     r^2/4 & \mbox{for
%     } r \ll 1. \\
%   \end{array} \right.
\end{split}
\end{equation}

\section{Adiabatic breakdown and optimal pumping frequency}\label{app:optimal}

\begin{figure}
   \includegraphics{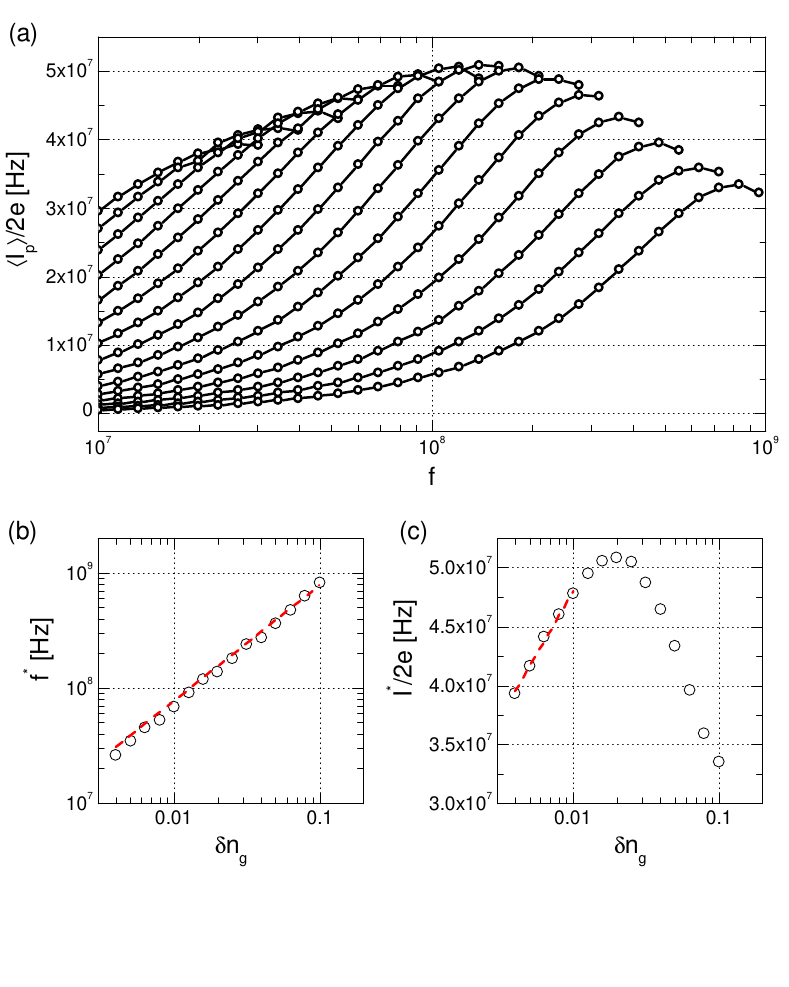}
   \caption{Optimal pumping frequency when $\varphi=\pi$.
   (a) Average pumped current
   $\mean{I_p}$ versus pumping frequency $f$ with $\varphi=\pi$ and $\dng$ taking a set of values
   in the range of 0.005 and 0.1 (from left to right).
   The other parameters are the same as in \figref{fig:nonadi}.
   (b,c) Optimal pumping frequency $f^*$ (b) and corresponding maximum pumped
   current $I^*$ (c) versus $\dng$.
   The dashed lines in (b) and (c) are linear fits to the numeric data close to $\dng=0$. 
   }
	\label{fig:optimal}
\end{figure}

Here we present a numeric optimization of the average pumped current $\mean{I_p}$ for the
case $\varphi=\pi$.

In \figref{fig:optimal}(a) we plot $\mean{I_p}$ versus $f$ for different values
of $\dng$ in the range of 0.005 and 0.1 (from left to right). The fact that each curve
reaches a maximum indicates that there is a tradeoff between speed-up gain and adiabaticity loss.
For each value of $\dng$, the optimal frequency $f^*(\dng)$ and the
corresponding maximum current $I^*(\dng)$ are plotted in Figs.~\ref{fig:optimal}(b) and
 \ref{fig:optimal}(c), respectively.
 From \figref{fig:optimal}(b) we see that the optimal frequency is approximately proportional to $\dng$.
 From \figref{fig:optimal}(c), that $I^*$ attains its maximum at a finite $\dng$.
 In particular, this shows that the optimal operation point is not arbitrarily close to $\dng=0$,
 as one might erroneously guess before taking nonadiabatic corrections into account.
 The scaling of both $f^*$ and $I^*$ is approximately linear as $\dng\to0$, as demonstrated by
 the linear fits shown as dashed lines in panels (b,c).
 
%  The scaling of $f^*$ and $I^*$ for $\dng\to0$ can be understood analytically.
%  In Sec.~\ref{sec:pumping:phi=pi} we have seen that $Q_p$ diverges as $1/\dng$
% for $\dng \to 0$ [Eq.~\eqref{eq:Qp_I_phi_pi}], while at the same time
% $\rho_{ge}$ diverges as $1/\dng^2T$ [Eq.~\eqref{eq:maxrhoge_phi=pi}].
% Putting things together, this implies that $I^*(\dng) \propto \dng$.

\clearpage

% \bibliography{FluxTh_biblio}

%merlin.mbs apsrev4-1.bst 2010-07-25 4.21a (PWD, AO, DPC) hacked
%Control: key (0)
%Control: author (8) initials jnrlst
%Control: editor formatted (1) identically to author
%Control: production of article title (-1) disabled
%Control: page (0) single
%Control: year (1) truncated
%Control: production of eprint (0) enabled
%

\end{document}